\documentclass{svproc}
\usepackage[T1]{fontenc}
\usepackage[utf8]{inputenc}
\usepackage{newunicodechar}
\newunicodechar{⋆}{\ensuremath{\star}}

\usepackage{graphicx}
\usepackage{amsmath,amssymb,amsfonts}
\usepackage{mathrsfs}
\usepackage[table]{xcolor}
\usepackage{booktabs}
\usepackage{multirow}
\usepackage{tabularx}
\usepackage{array}
\usepackage{makecell}
\usepackage{float}
\usepackage{caption}
\usepackage{subcaption}
\usepackage{enumitem}
\usepackage{ragged2e}
\usepackage[most]{tcolorbox}
\usepackage{listings}
\usepackage{textcomp}
\usepackage[title]{appendix}
\usepackage{hyperref}
\usepackage{xcolor}
\usepackage{mdframed}

\newmdenv[
  backgroundcolor=gray!10,
  linecolor=black!25,
  linewidth=0.6pt,
  roundcorner=0pt,
  innerleftmargin=8pt,
  innerrightmargin=8pt,
  innertopmargin=6pt,
  innerbottommargin=6pt,
  skipabove=6pt,
  skipbelow=6pt
]{rqbox}

\tcbuselibrary{listingsutf8}
\newtcblisting{promptlisting}{
  listing only,
  colback=gray!5,
  colframe=black!50,
  listing options={
    basicstyle=\ttfamily\footnotesize,
    breaklines=true,
    columns=fullflexible
  },
  left=0pt, right=0pt, top=2pt, bottom=2pt,
  boxsep=2pt,
}

\lstdefinestyle{jsonstyle}{
    basicstyle=\ttfamily\footnotesize,
    backgroundcolor=\color{gray!10},
    frame=single,
    breaklines=true,
    showstringspaces=false,
    literate=
     *{0}{{{\color{blue}0}}}{1}
      {1}{{{\color{blue}1}}}{1}
      {2}{{{\color{blue}2}}}{1}
      {3}{{{\color{blue}3}}}{1}
      {4}{{{\color{blue}4}}}{1}
      {5}{{{\color{blue}5}}}{1}
      {6}{{{\color{blue}6}}}{1}
      {7}{{{\color{blue}7}}}{1}
      {8}{{{\color{blue}8}}}{1}
      {9}{{{\color{blue}9}}}{1}
      {:}{{{\color{red}:}}}{1}
      {,}{{{\color{red},}}}{1}
      {"}{{{\color{orange}"}}}{1}
}

\begin{document}
\mainmatter              
\title{Adopt-an-AI-Researcher: Bootstrapping Industry-Academia Collaboration to Identify and Prioritise Generative AI Use Cases}
\titlerunning{Adopt-an-AI-Researcher}  
%
\author{
Malik Abdul Sami\inst{1}\thanks{Corresponding author.} \and
Zeeshan Rasheed\inst{1}\thanks{Corresponding author.} \and
Meri Olenius\inst{1} \and
Muhammad Waseem\inst{1} \and
Kai-Kristian Kemell\inst{1} \and
Jussi Rasku\inst{1} \and
Pekka Abrahamsson\inst{1}
}

\authorrunning{Sami et al.}

\institute{
Faculty of Information Technology and Communication Sciences,\\
Tampere University, Tampere, Finland\\
\email{
malik.sami@tuni.fi,
zeeshan.rasheed@tuni.fi,
meri.olenius@tuni.fi,
muhammad.waseem@tuni.fi,
kai-kristian.kemell@tuni.fi,
jussi.rasku@tuni.fi,
pekka.abrahamsson@tuni.fi
}
}

\maketitle              

\begin{abstract}
Organisations want to use generative AI (GenAI), but many do not know where
to begin. Researchers understand AI capabilities, while employees understand
their organisation's processes, data, constraints, and business needs, and
effective adoption requires both forms of knowledge. We introduce
\textit{Adopt-an-AI-Researcher}, an approach that brings them together. AI
researchers work across organisational functions for a fixed period to
identify adoption needs, develop AI use cases, and support their
prioritisation. We applied and examined the approach through a six-week case
study in a mid-sized European energy company. Data were collected through 16
semi-structured group interviews across nine organisational functions,
observations, and internal documents, and were analysed iteratively. The
analysis identified five themes: manual and repetitive work, forecasting and
predictive analytics, data fragmentation and integration, compliance and
validation, and organisational and infrastructure readiness. The engagement
produced 41 AI use cases, a use case taxonomy, prioritised use cases, and two
proof-of-concept demonstrations. The case provides initial evidence that the
approach is feasible and can move an organisation from a general interest in
AI to documented adoption needs, structured use cases, and priorities for
further assessment. It also shows that data readiness, system integration,
validation requirements, and human review shape which AI use cases can be
pursued.

\end{abstract}

\section{Introduction}
\label{sec:introduction}
Generative AI (GenAI) capabilities are advancing faster than many
organisations can adopt them \cite{kemell2025still}. AI researchers understand
current AI capabilities, limitations, and technical requirements, while
employees understand the organisation's processes, data, constraints, and
business needs \cite{marijan2022industry}. Effective AI adoption requires these
two forms of knowledge to meet. When they remain separate, organisations
struggle to identify where AI can provide value and which conditions are
required for successful adoption
\cite{HOSSEINI2025100399,sami2026bridging}.

Research on industry--academia collaboration shows that knowledge transfer is
most effective when researchers and practitioners work together on shared
problems \cite{marijan2022industry}. Previous studies describe continuous
co-experimentation, controlled industrial case studies, and long-term
collaboration between researchers and companies
\cite{MIKKONEN201834,salo2004empirical}. Other studies identify management
support, shared goals, organisational access, and communication as important
conditions for successful collaboration
\cite{wohlin2012success,garousi2016challenges,garousi2019characterizing}.

However, these studies mainly explain collaboration after a research
partnership and a shared problem already exist. Many organisations are
interested in AI but have no research partner, no defined AI project, and no
structured way to connect AI capabilities with business needs. As a result,
identifying AI use cases, assessing adoption needs, and deciding where to begin
remain difficult. This challenge is particularly pronounced in organisations
that rely on a combination of in-house systems, cloud services, purchased
software, and third-party integrations, but lack internal AI expertise and a
structured starting point for AI adoption.

We address this problem through the \textit{Adopt-an-AI-Researcher} approach,
in which an organisation hosts AI researchers across its organisational
functions for an agreed period. The approach combines researchers' knowledge
of AI capabilities with employees' knowledge of processes, data, constraints,
and business needs. It starts from organisational problems rather than
predefined technologies and iteratively identifies and refines AI use cases
through interviews, analysis, stakeholder validation, and reporting. The
engagement has a fixed duration but a flexible scope. The outcome is a
documented set of adoption needs, candidate use cases, and priorities that
guide the organisation's first AI initiatives.

We applied this approach through a six-week case study in a mid-sized European
energy company involving nine organisational functions and 16 semi-structured
group interviews. The study examined the feasibility of the approach and
identified AI adoption needs, potential use cases, and lessons from its
application. We address the following research questions:

\begin{rqbox}
\noindent\textbf{RQ1:} How can the Adopt-an-AI-Researcher approach be applied
in an organisation at the start of AI adoption, and what outcomes does it
produce?
\end{rqbox}

\begin{rqbox}
\noindent\textbf{RQ2:} What AI adoption needs, use cases, and lessons emerge
when the approach is applied in an industrial organisation?
\end{rqbox}

\noindent\textbf{Contributions.}
This paper makes two contributions. First, it introduces the
\textit{Adopt-an-AI-Researcher} approach, defining the organisational entry
point, required roles, preparation activities, iterative analysis, stakeholder
validation, and reporting. Second, it applies and examines the approach
through a six-week industrial case study and reports AI adoption needs, 41 AI
use cases, a use case taxonomy, prioritised use cases, and implications for
research and practice.

\noindent\textbf{Structure of the paper.} Section~\ref{sec:related_work}
presents related work. Section~\ref{sec:approach} introduces the proposed
approach. Section~\ref{sec:case} describes the case study.
Section~\ref{sec:results} presents the results. Section~\ref{sec:discussion}
discusses the findings, implications, and threats to validity. Finally,
Section~\ref{sec:conclusion} concludes the paper and outlines future work.

\section{Related Work}
\label{sec:related_work}

\textbf{Organising industry--academia collaboration.}
Mikkonen \textit{et al.}~\cite{MIKKONEN201834} propose continuous
co-experimentation, in which companies and research institutes jointly work
on shared problems through rapid and iterative cycles. Salo and
Abrahamsson~\cite{salo2004empirical} present a controlled industrial case
study approach with a fixed implementation period, flexible scope, and both
industrial and scientific outcomes. Wohlin \textit{et
al.}~\cite{wohlin2012success} identify factors supporting successful
industry--academia collaboration, while Garousi \textit{et
al.}~\cite{garousi2016challenges,garousi2019characterizing} synthesise common
challenges and collaboration patterns. These studies mainly
address collaborations in which the partners, the research setting, or the
method under study are determined in advance, providing limited guidance for
organisations without a research partner or a defined AI problem.

\textbf{GenAI use at the task and team levels.}
Prior work has mainly examined how practitioners use GenAI tools for specific tasks. Liang \textit{et al.}~\cite{liang2024large} found that developers mainly use these tools for syntax recall, reducing manual typing, and completing tasks faster, while also reporting limited user control and recommendations that do not always meet practical needs. Davila \textit{et al.}~\cite{davila2024industry} reported limited context awareness in tool suggestions, while Ulfsnes \textit{et al.}~\cite{ulfsnes2024transforming} found that developers sometimes turn to GenAI instead of colleagues, changing communication and learning within teams. These studies explain task-level use but not how organisations identify where AI should be introduced across organisational functions. \textbf{Organisational determinants of AI adoption.}
Research on AI adoption identifies workflow compatibility, leadership support, organisational readiness, use-case clarity, data protection, and security as important factors~\cite{russo2024navigating,polisetty2024determines,solaimani2024exploration,hana2026ai,brehme2025retrieval}. Russo~\cite{russo2024navigating} further reports that early adoption depends strongly on how well AI tools fit existing workflows. J\"ohnk \textit{et al.}~\cite{johnk2021ready} derived five categories of organisational AI readiness factors, together with indicators for assessing them, from an interview study with AI experts. These studies explain which factors affect adoption but provide less detail on how they are examined across organisational functions while AI use cases are identified and prioritised. \textbf{Studies identifying AI use cases.}

Kemell \textit{et al.}~\cite{kemell2025still} conducted a multiple-case study of seven European software organisations and identified 25 use cases together with adoption challenges, including data privacy, legislation, and change resistance. Flyckt \textit{et al.}~\cite{flyckt2025identifying} identified and prioritised AI tasks in make-to-order manufacturing and highlighted data requirements and explainability. In the energy sector, Pfeifer \textit{et al.}~\cite{pfeifer2024use} examined how AI can be applied in utility companies and how it affects innovation, but did not study how use cases are identified and prioritised across organisational functions.

The reviewed work assumes that earlier steps have already been
taken. Continuous co-experimentation \cite{MIKKONEN201834} is carried out
within established consortia and research programmes, where the collaborating
partners are already in place. The controlled case study approach
\cite{salo2004empirical} evaluates a development method chosen before the
study begins, within a single project team, and its outcome is a working
software system together with research data on that method. Adoption and
readiness studies \cite{russo2024navigating,johnk2021ready} report the factors
that determine whether an organisation is ready to adopt AI, but not the AI
applications it should select. Use-case studies
\cite{kemell2025still,flyckt2025identifying} report the use cases found, but
not how a company without a research partner or in-house AI expertise can find
them.

We propose \textit{Adopt-an-AI-Researcher} for this earlier
step. It starts without an existing research partnership, a defined AI
project, or a predetermined technology, and without in-house AI expertise to
define one. Researchers work across every organisational function rather than
on a single selected problem, so use cases from different functions can be
compared against each other. The result is a ranked list of candidate use
cases together with the data, system, and validation conditions each one
depends on. This gives existing collaboration models the defined problem they
take as their starting point. We report what the approach produced in a
mid-sized energy company across nine organisational functions, where
the state of the data infrastructure shaped which use cases could be pursued.

\section{The Adopt-an-AI-Researcher Approach}
\label{sec:approach}

\begin{figure}[t]
    \centering
    \includegraphics[width=1.05\textwidth]{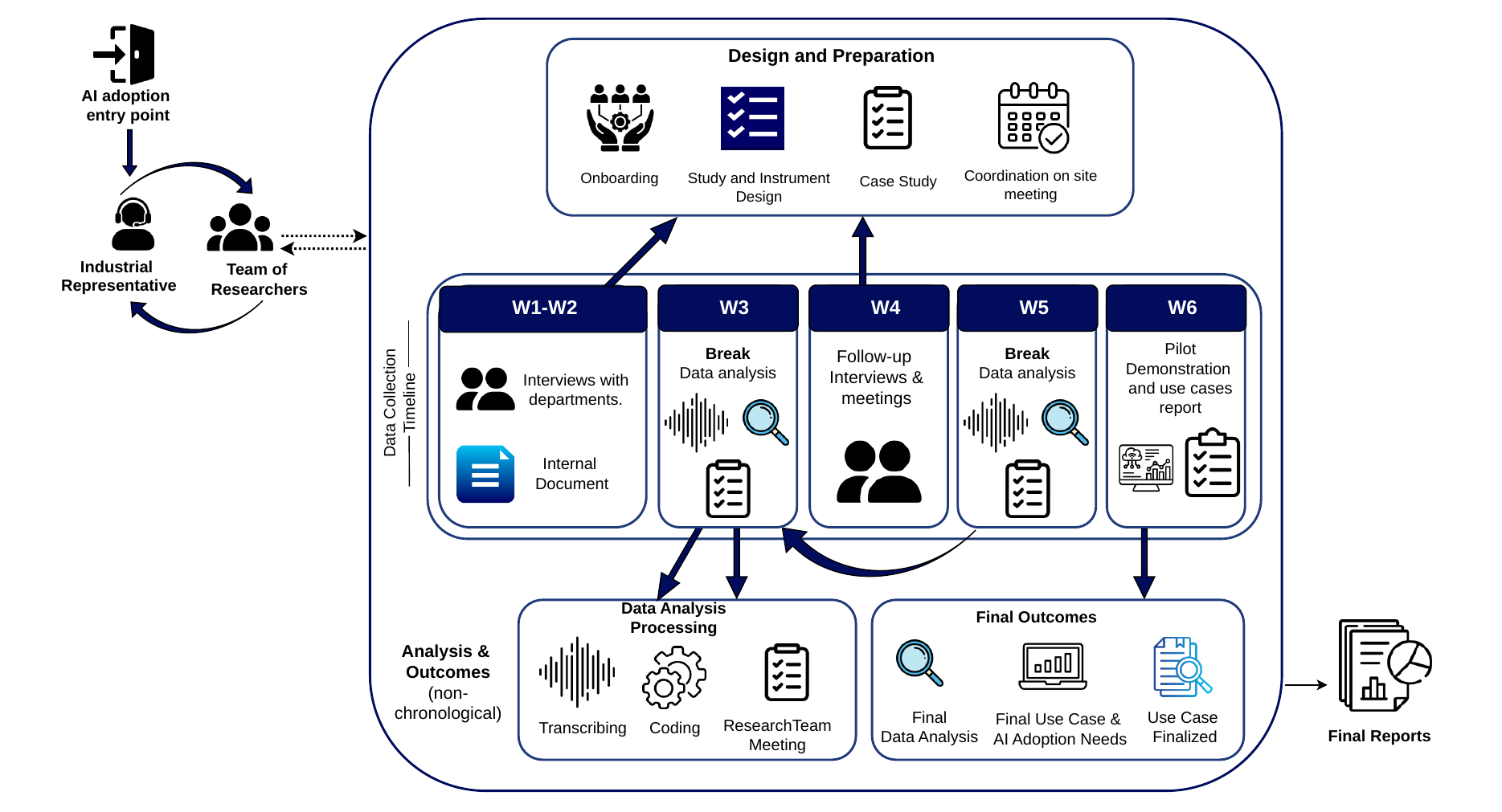}
    \caption{The Adopt-an-AI-Researcher Approach.}
    \label{fig:study_workflow}
\end{figure}

We use AI as the umbrella term throughout this paper; GenAI refers to
generative techniques such as large language models, one subset of the AI
methods discussed. We propose \textit{Adopt-an-AI-Researcher}, a time-boxed
industry--academia collaboration approach for organisations that want to
explore AI adoption but do not yet have a clearly defined AI project or an
established research collaboration. The organisation hosts AI researchers
for an agreed period to examine organisational challenges, identify possible
AI applications, and clarify the conditions required for adoption. The approach starts from problems identified by the organisation rather than from technologies or solutions selected in advance. Its duration is fixed,
while the scope can be adjusted as organisational needs become clearer.
Data collection and analysis are conducted iteratively so that emerging findings can guide follow-up activities \cite{MIKKONEN201834}.
Figure~\ref{fig:study_workflow} presents the \textit{Adopt-an-AI-Researcher} approach
and its application in the industrial case study.

\textbf{AI adoption entry point.}
The approach begins when an organisation decides to explore AI adoption
without committing to a predefined project or technical solution. The
organisation identifies its motivation, such as reducing manual work,
improving decision-making, addressing data-related challenges, or examining
AI opportunities across organisational functions.

\textbf{Roles, design, and preparation.}
The organisation appoints an industrial representative who acts as the main
contact between the organisation and the research team. The representative
identifies relevant organisational functions and participants, coordinates
activities, supports access to internal material, and explains
organisation-specific terminology and practices.
The AI researchers design and conduct the engagement. Their responsibilities
include preparing the study instruments, facilitating interviews and
meetings, documenting observations, analysing the material, identifying AI
use cases, and preparing recommendations. A supervising researcher supports
the research design and reviews the analysis.
Preparation includes researcher onboarding, clarification of the
organisation's motivation and scope, participant selection, activity
coordination, and preparation of the interview guide, observation structure,
and document review process.

\textbf{Data collection timeline.}
The researchers collect data from relevant organisational functions through
 interviews, internal documents, process descriptions,
observations, meetings, and informal discussions. Data collection examines
current work practices, operational problems, time-consuming activities,
existing workarounds, data availability, AI expectations, and adoption
concerns. The engagement follows an agreed timeline. However, its exact duration,
number of participants, and number of activities may differ between
organisations.

\textbf{Iterative analysis.}
Analysis begins during data collection rather than after all organisational
activities have finished. The researchers use transcription, meeting notes,
post-session summaries, coding, memoing, and research-team discussions to
identify emerging findings.
These findings inform follow-up questions, clarify assumptions, and support
the refinement of candidate AI use cases. This creates an iterative connection
between data collection, analysis, and organisational follow-up.

\textbf{Validation and reporting.}
The researchers consolidate the identified organisational challenges, AI use
cases, and adoption needs. These are presented to relevant organisational
stakeholders for review, prioritisation, and validation.
Stakeholder feedback is used to refine the use cases and clarify their
business relevance, feasibility, expected value, and organisational
requirements. The engagement concludes with a report containing the final
use cases, adoption needs, priorities, and recommendations for further work.
A pilot or demonstration may also be included when suitable, but it is not a
required part of the approach.

\section{Case Study}
\label{sec:case}

We applied the \textit{Adopt-an-AI-Researcher} approach in a single embedded case study~\cite{yin2017case}. The study is exploratory and served two purposes: first, to examine how our proposed approach works in a case company and, second, to produce empirical findings on organisational AI adoption needs and potential AI use cases.

\textbf{Case context.}
The case organisation was a mid-sized European energy company operating across energy trading, heating, solar, wind, and related services. The organisation integrated more than 100 information systems and platforms, served thousands of customers, and relied on distributed cloud-based digital infrastructure. The engagement was initiated by the company. Management wanted to understand which AI applications were feasible across its organisational functions, which technical, organisational, and regulatory conditions could affect their adoption, how productivity could be increased, and which daily routine tasks or pain points could be addressed using GenAI. The company appointed a senior representative with visibility across all its organisational functions and units. The representative coordinated access, identified participants, scheduled sessions, provided access to internal material, and clarified domain-specific terminology. Two AI researchers conducted the interviews and attended sessions, supported by a supervising researcher.

Nine organisational functions participated, with management represented within these functions rather than treated as a separate organisational function.

Each organisational function is referred to using the abbreviation OF (organisational function) followed by a number, e.g., OF1. Table~\ref{tab:case_functions} presents the organisational functions and their interview coverage.

\begin{table}[t]
\centering
\caption{AI use cases identified per organisational function}
\label{tab:case_functions}
\scriptsize
\begin{tabularx}{\textwidth}{lXc}
\toprule
\textbf{Code} & \textbf{Organisational function} & \textbf{Use cases} \\
\midrule
OF1 & Customer-facing operational processes & 4 \\
OF2 & Core operational and infrastructure processes & 4 \\
OF3 & Information systems and data integration & 7 \\
OF4 & Financial planning and reporting processes & 5 \\
OF5 & Market-oriented operational processes & 6 \\
OF6 & Asset- and risk-related planning activities & 6 \\
OF7 & Workforce planning and internal capability development & 3 \\
OF8 & Strategic project and initiative development & 5 \\
OF9 & Business development and feasibility analysis & 1 \\
\midrule
\textbf{Total} & & \textbf{41} \\
\bottomrule
\end{tabularx}
\end{table}

\textbf{Data collection and timeline.}
The engagement ran from March to April 2025 over six weeks. Four weeks included direct activities with the organisation, while two intervening weeks were dedicated to off-site analysis. The timeline is shown in Figure~\ref{fig:study_workflow}. Two researchers worked approximately six hours per day. The organisation contributed approximately 24 hours of group interview sessions, excluding coordination and follow-up activities.

During Weeks~1 and~2, the researchers completed onboarding, coordinated the study with the company representative, reviewed internal material, and conducted initial interviews across the nine organisational functions. The sessions examined participants' responsibilities, current processes, time-consuming activities, existing workarounds, data and system dependencies, forecasting and decision-making practices, possible AI applications, and adoption concerns. During Week~3, the researchers conducted the first off-site analysis by transcribing available recordings, reviewing meeting notes and internal documents, preparing post-session summaries, conducting initial coding, and discussing emerging findings within the research team. The analysis identified missing information and informed the preparation of the follow-up sessions.

During Week~4, the researchers returned for follow-up interviews and meetings to clarify earlier findings, examine emerging AI opportunities, and validate their interpretation of the identified organisational challenges. During Week~5, the researchers consolidated overlapping findings, refined the candidate use cases, examined related adoption needs, and prepared the findings for organisational review. During Week~6, the researchers presented the proposed use cases and their prioritisation to company stakeholders. Stakeholder feedback informed the final refinement of the use cases and preparation of the industrial report. The report was submitted at the end of April 2025, and the findings were presented to the company in May 2025.

The primary data consisted of 16 semi-structured group interview sessions across nine organisational functions. Participants were selected purposively based on their knowledge of organisational workflows, operational challenges, and potential AI adoption opportunities. Senior employees and heads of organisational functions were prioritised because they could provide an overview of both function-specific and cross-functional processes. Participants represented technical, operational, financial, managerial, strategic, and governance roles, with industry experience ranging from four to more than fifteen years. Fifteen employees participated across the sixteen sessions, distributed across all nine organisational functions; some attended both the initial and the follow-up session for their function. Table~\ref{tab:profiles} summarises their role clusters.

\begin{table}[t]
\centering
\caption{Participant role clusters}
\label{tab:profiles}
\scriptsize
\begin{tabularx}{\textwidth}{Xc}
\toprule
\textbf{Role cluster} & \textbf{N} \\
\midrule
Technical engineering roles        & 4 \\
Customer-facing operational roles  & 3 \\
Market and finance roles           & 3 \\
Managerial and strategic roles     & 3 \\
Support and governance roles       & 2 \\
\midrule
\textbf{Total} & \textbf{15} \\
\bottomrule
\end{tabularx}
\end{table}

Group size ranged from two to five participants. Seven organisational functions participated in two sessions, while two participated in one session. Most sessions lasted approximately 1.5 hours, resulting in approximately 24 hours of interview data. Interviews were conducted mainly on-site, although some participants joined online. Group interviews were selected to capture shared understandings of work practices and cross-functional views of AI adoption. The semi-structured format allowed follow-up questions based on earlier sessions and emerging findings. The sessions were guided by the questions in Table~\ref{tab:interview_guide}, which moved from current practices and data dependencies towards candidate AI applications and adoption concerns.

\begin{table}[t]
\centering
\caption{Guiding questions for the semi-structured group interviews}
\label{tab:interview_guide}
\scriptsize
\setlength{\tabcolsep}{3pt}
\renewcommand{\arraystretch}{1.1}
\begin{tabularx}{\textwidth}{p{0.15\textwidth} X}
\toprule
\textbf{Topic} & \textbf{Question} \\
\midrule
Context & What is your role and responsibility at the company? \\
Context & Can you briefly explain how your organisational function operates? \\
\midrule
Current work & What are the main challenges your function faces in day-to-day operations? \\
Current work & Which tasks or processes consume the most time or resources? \\
Current work & How are these challenges currently addressed (manual methods, tools, workarounds)? \\
\midrule
Data and systems & Which systems and data sources do you use in your daily work? \\
Data and systems & Where does information have to be moved or re-entered manually between systems? \\
Data and systems & How do you currently make predictions, forecasts, or decisions in your work? \\
\midrule
AI opportunities & Where do you see opportunities for AI or automation to improve efficiency? \\
AI opportunities & Are there areas where errors often occur that AI could help reduce? \\
AI opportunities & How should AI outputs be presented so they are actionable for your team? \\
AI opportunities & Which challenges in your function also occur across other functions? \\
\midrule
Priorities and concerns & Which of these challenges should be prioritised for AI-driven solutions? \\
Priorities and concerns & What concerns or hesitations do you have about AI adoption? \\
\bottomrule
\end{tabularx}
\end{table}

\textbf{Observations, notes, and supporting documents.} During the on-site
weeks, the researchers observed meetings, informal discussions, and team
interactions. The researchers recorded field notes using a meeting template
that captured the systems mentioned, the workarounds described, and points
where participants disagreed or added to each other's accounts. For the eight
interviews that were not audio-recorded, these notes were the main record and
were written up immediately afterwards. The company representative also
provided internal documents, process descriptions, and system outlines, which
helped the researchers understand the interview material in context. After
each interview, the researchers prepared a short summary of the operational
problems raised, the data and system dependencies mentioned, and the possible
AI applications discussed. These summaries served as the working record for
drafting and refining the candidate use cases in the following weeks.

\textbf{Data analysis.}
We analysed the qualitative material using an inductive thematic coding approach~\cite{braun2006using}. Analysis took place on two timescales. First, rapid analysis was conducted throughout the six-week engagement using post-session summaries, initial coding, document review, and research-team discussions. The resulting findings guided follow-up questions, refined candidate AI use cases, and supported organisational validation. Second, a detailed thematic analysis was conducted after the industrial report was submitted.

The eight transcribed interviews formed the primary material for detailed coding. The first author iteratively coded the transcripts and grouped related codes into higher-level categories. Coding decisions and emerging categories were discussed with the other authors. The detailed coding produced 166 coded quotations. After overlapping codes were consolidated, the dataset contained 125 unique codes organised into 14 code groups and five higher-level themes. Although detailed thematic coding focused on the eight transcripts, all 16 interviews, meeting notes, observations, internal documents, and post-session summaries informed the identification and refinement of the AI use cases.

{\textbf{Ethics}}: Participation was voluntary, with no additional compensation. Before each interview, participants received information about the study purpose and data handling and provided informed consent.

\section{Results}
\label{sec:results}

The results are presented according to the two research questions. For RQ1,
we report what resulted from applying the \textit{Adopt-an-AI-Researcher} approach in
the case organisation. For RQ2, we report the AI adoption needs identified
through thematic analysis, the resulting use cases, their prioritisation, and
the proof-of-concept demonstrations.

\subsection{RQ1: Application of the Adopt-an-AI-Researcher Approach}

The \textit{Adopt-an-AI-Researcher} approach was applied over the planned six-week
period. The industrial representative provided access to nine organisational
functions, identified relevant participants, coordinated the sessions, and
supported access to internal information. The alternating periods of
organisational engagement and analysis allowed the researchers to examine the
initial findings before conducting follow-up sessions.

The engagement produced 41 organisational-function-specific AI use cases, a
prioritised list of use-case categories, an industry report, and two initial
proof-of-concept demonstrations. The findings and use cases were presented to
company stakeholders for review. These outcomes provide initial evidence that
the approach can be applied within a bounded industrial engagement. However,
the case does not establish its long-term effectiveness or applicability in
other organisations.

\subsection{RQ2: AI Adoption Needs and Use Cases}

The thematic analysis identified five themes, summarised in
Table~\ref{tab:themes}.

\begin{table}[t]
\centering
\caption{High-level themes and their code groups from the interviews.}
\label{tab:themes}
\scriptsize
\setlength{\tabcolsep}{3pt}
\renewcommand{\arraystretch}{0.95}
\begin{tabularx}{\textwidth}{X X r}
\toprule
\textbf{High-level theme} & \textbf{Code groups} & \textbf{Total} \\
\midrule
Manual and Repetitive Work &
Manual and repeated work (12), work management (11), reporting (6) & 29 \\
Forecasting and Predictive Analytics &
Forecasting (12), predictive maintenance (10), investment planning (5),
market analysis (4) & 31 \\
Data Fragmentation and Integration &
Data integration (9), data quality (9), system infrastructure (12) & 30 \\
Compliance and Validation &
Data validation issues (8), compliance and financial processes (7),
vendor management (6) & 21 \\
Organisational and Infrastructure Readiness &
Asset management (4), other operational challenges (10) & 14 \\
\midrule
\textbf{Total} & \textbf{14 code groups} & \textbf{125} \\
\bottomrule
\end{tabularx}
\end{table}

\textbf{Theme 1: Manual and Repetitive Work.}
Employees described routine activities that require repeated checks,
corrections, and information validation. These tasks take time and affect
daily productivity.

Participants linked AI to routine tasks such as reporting, validation, and
repeated data checks. A participant from OF4 described a possible role for
automation in checking processes, noting, \textit{``AI could do checking for us and we then approved''} (OF4). This shows that employees identify repetitive reporting and
checking tasks as suitable for automation. At the same time, uncertainty remains about the extent of AI support in these processes.

\textbf{Theme 2: Forecasting and Predictive Analytics.}
Forecasting was among the most frequently raised themes and
appeared across several organisational functions.

For instance, a participant from OF9 described forecasting practices in long-term planning as follows: \textit{``We make a plan for 2036. We try to
forecast as exact as we can how many customers we will have and what kind of
techniques we will use in the future, for example dismantling overhead lines
and replacing them with underground cables''}. Similarly, a participant from OF2 described forecasting practices related to changes in the operating environment as follows: \textit{``First one is like a strategic forecast of changes in the
operating environment in the electricity distribution network. So it is like
forecasting thing. Different kind of things that we forecast. What is changing
in our field here''}.

These statements show that participants linked AI to forecasting accuracy,
predictive maintenance, scenario planning, and long-term decision-making.

\textbf{Theme 3: Data Fragmentation and Integration.}
Participants described disconnected systems, inconsistent data, and limited
automation in monitoring processes. These issues increase manual effort and
affect the reliability of information across organisational functions. For instance, a participant from OF3 described the workload caused
by manual monitoring, noting, \textit{``We have error handling, but many
processes we do not have because the amount of work is so big. It could be
better if some system monitored and categorized the errors automatically''}.

This shows that participants expect automation to support system
integration and improve data quality. They also indicate that data access,
integration, infrastructure, and automated monitoring are requirements for
several proposed AI applications.

\textbf{Theme 4: Compliance and Validation.}
A participant from OF4 expressed interest in automated validation,
explaining, \textit{``I would like to see some kind at least reading… AI to
train and detect''}. Another participant from OF2 discussed the
potential of anomaly detection, noting, \textit{``If we find anomalies, yeah,
predict anomalies. So it can also help in that, not just to find, but to
detect failures of the network by using GenAI. It affects a lot, also the
prices that we can take from our customers''}.

These findings show that participants associate AI with validation, anomaly
detection, and support for compliance-related work. Participants mainly
described AI as support for checking and decision-making, particularly where
errors could affect operations, customers, or financial processes.

\textbf{Theme 5: Organisational and Infrastructure Readiness.}
Readiness-related statements focused on AI as a supportive tool that could
assist with tasks without immediately replacing existing systems. A
participant from OF4 suggested that AI systems should operate incrementally,
noting, \textit{``Only just suggest you those issues or it would be easier if
system assisted''}. A participant from OF6 referred to workload
concerns, explaining, \textit{``There will not be more people to do this,
maybe less. So anything manual or time consuming that we can reduce helps the
workload''}.

These findings show that participants prefer gradual integration of AI into
existing workflows. Operational constraints and limited resources shaped
adoption decisions, while human review was preferred over fully automated
operation.

\textbf{AI use cases and prioritisation.}
The thematic findings and the rapid analysis conducted during the engagement
informed the identification of 41 organisational-function-specific AI use
cases, listed in Table~\ref{tab:use_cases} by category in
priority order. Each use case is assigned to exactly one of six categories.
The first number in each use-case identifier denotes the organisational
function, while the second identifies the use case within that function.

\begin{table}[!t]
\centering
\caption{The 41 AI use cases, grouped by category in priority order}
\label{tab:use_cases}
\scriptsize
\setlength{\tabcolsep}{4pt}
\renewcommand{\arraystretch}{1.0}
\begin{tabularx}{\textwidth}{c p{0.26\textwidth} l X}
\toprule
\textbf{Priority} & \textbf{Category} & \textbf{ID} & \textbf{Use case} \\
\midrule
1 & Reporting & 2.1 & Component reporting \\
  &           & 2.3 & Network reporting \\
  &           & 2.4 & Development planning reports \\
  &           & 4.3 & Financial insights \\
  &           & 4.5 & Financial report \\
  &           & 5.4 & Trading report \\
  &           & 5.5 & Reporting accuracy monitoring \\
  &           & 6.4 & Risk assessment and reporting \\
  &           & 6.5 & Report review \\
  &           & 7.2 & Employee survey analysis \\
\midrule
2 & RAG and agentic RAG-based solutions & 1.1 & Rule-based chatbots \\
  &           & 1.2 & Feedback reviews \\
  &           & 1.3 & Email handling \\
  &           & 1.4 & Phone calls \\
  &           & 4.1 & Data retrieval \\
  &           & 5.2 & Email automation \\
  &           & 5.3 & Email and file search \\
  &           & 6.6 & Feedback and follow-up \\
  &           & 7.1 & Sponsorship application processing \\
  &           & 7.3 & Internal queries chatbot \\
\midrule
3 & Predictive maintenance & 2.2 & Fault detection in distribution lines \\
  &           & 8.2 & Failure prediction \\
\midrule
4 & Anomaly detection & 3.2 & User activity monitoring \\
  &           & 3.4 & Data privacy risk analysis \\
  &           & 3.5 & Alerting system \\
  &           & 3.7 & Security monitoring \\
  &           & 4.4 & Fraud detection \\
  &           & 8.1 & Safety data monitoring \\
\midrule
5 & Budgeting and forecasting & 3.6 & License cost simulation \\
  &           & 4.2 & Invoice validation \\
  &           & 5.1 & Trading forecast \\
  &           & 5.6 & Revenue planning \\
  &           & 6.2 & Budget planning forecast \\
  &           & 8.3 & Pricing forecast \\
  &           & 8.5 & Billing verification \\
\midrule
6 & Other & 3.1 & Integration map maintenance \\
  &           & 3.3 & Data ingestion platform \\
  &           & 6.1 & Process optimisation \\
  &           & 6.3 & Asset optimisation \\
  &           & 8.4 & Workforce allocation \\
  &           & 9.1 & Early-stage feasibility assessment \\
\bottomrule
\end{tabularx}
\end{table}

The six categories were ordered qualitatively by the research
team, without a numerical scoring model, based on relevance to the needs raised
across functions, ease of implementation, and expected organisational value.
Ease of implementation was judged mainly by whether a use case could draw on
information sources that already existed and whether its output had a clear
point of human review.

The order does not follow the frequency of the themes.
Forecasting was among the most frequently raised needs, yet budgeting and
forecasting ranked fifth, because these use cases depended on integrated
historical data that the organisation did not yet hold in a usable form.
Reporting and RAG-based retrieval ranked first and second because they relied
on documents and reports that already existed, and their outputs would be
reviewed by employees before use. Predictive maintenance ranked third despite
similar data requirements, because failures in the distribution network carry
high operational and financial cost, giving these use cases high expected
value. The priority order therefore reflects what could be started with the
organisation's current data and systems, weighed against the value of
addressing each need. The prioritised use cases were presented to company
stakeholders and included in the industrial report.

\textbf{Proof-of-concept demonstrations.}
Two use cases from the RAG-based category, which could be built
on existing documents and email records, were taken forward as proof-of-concept
demonstrations with the partner organisation: a RAG-based system that drafts
responses to incoming customer emails for review and approval by a human
operator, and a RAG-based document search system that answers natural-language
queries over internal documents and returns references to the source files.
Both were developed as demonstrations rather than production deployments,
and their evaluation in operational use is left for future work.

\section{Discussion}
\label{sec:discussion}

The \textit{Adopt-an-AI-Researcher} approach was applied in an
exploratory six-week case study in a mid-sized energy company involving nine
organisational functions. This section discusses the findings in relation to
the research questions and their implications for research and practice.

\subsection{RQ1: The Adopt-an-AI-Researcher Approach as an Organisational Entry Point}

The case provides initial evidence that the proposed approach can be carried
out at the beginning of AI adoption, before a defined AI project or an
established research collaboration exists. Unlike continuous
co-experimentation \cite{MIKKONEN201834}, it did not depend on an established
consortium, and unlike the controlled case study approach
\cite{salo2004empirical}, it did not start from a method selected in advance.
In the case, this allowed the engagement to cover all nine organisational
functions and to identify use cases that had not previously been articulated.

Many organisations want to adopt AI but do not know where to start, and
researchers without a partnership have no route to the problems in
organisations where AI could help. The \textit{Adopt-an-AI-Researcher}
approach addresses this point. The organisation hosts AI researchers for a
fixed period to identify where AI can be applied, and the researchers gain
access to the problems employees face in their work. In the case, the
engagement started from employees' pain points, including tasks that take
time, checks done repeatedly, and workarounds, rather than from AI
capabilities, and moved from these towards the places where AI could be used.
This gave the organisation, at the start of its AI adoption, a set of use
cases grounded in its own processes and data, and gave the researchers access
to problems and data dependencies across nine organisational functions.

Three elements supported the engagement. First, the industrial representative
coordinated access across organisational functions, identified participants,
and supported communication between the company and the research team, which
made it possible to cover nine functions within four on-site weeks. Second,
alternating data collection and analysis allowed findings to guide follow-up
sessions and the refinement of the use cases: the Week~3 analysis identified
missing information, which the Week~4 follow-up sessions then addressed.
Third, the fixed duration and flexible scope allowed the organisation to
explore AI opportunities without committing to a predefined solution. These
elements helped move the organisation from an interest in AI to a set of
adoption needs and use cases in priority order.

\subsection{RQ2: Organisational AI Adoption Needs and Use Cases}

The findings show that the main task was not only to generate candidate AI use
cases. The organisation also needed to determine which opportunities could be
pursued within its existing data, systems, resources, and organisational
conditions. The engagement produced more candidate use cases than could be
addressed immediately. Prioritisation was therefore used to distinguish
between opportunities that could be examined first and those that required
further preparation.

Comparing the themes with the priority order shows that the needs raised most
often were not those that could be started first. In this case, data
readiness rather than demand determined where AI adoption could begin. This is
consistent with prior research that identifies organisational readiness and
data-related conditions as factors in AI adoption
\cite{polisetty2024determines,solaimani2024exploration,brehme2025retrieval},
and with the view of J\"ohnk \textit{et al.}~\cite{johnk2021ready} that
readiness should be assessed for a specific adoption purpose. The case shows
what this means in practice: the same organisation was ready for reporting and
retrieval use cases, which drew on existing documents, but not yet for
forecasting use cases, which required integrated historical data.

These findings relate to two studies discussed in
Section~\ref{sec:related_work}. Kemell \textit{et al.}~\cite{kemell2025still}
identified 25 use cases across seven software organisations and reported data
privacy, legislation, and change resistance as adoption challenges. Our case
extends this picture to operational, financial, customer-facing, and
infrastructure functions outside software development, where a recurring
constraint was the fragmentation of data across more than 100 integrated
systems. This constraint affected not only whether use cases were feasible,
but also the order in which they could be pursued. Flyckt
\textit{et al.}~\cite{flyckt2025identifying} highlighted data requirements and
explainability when prioritising AI tasks in manufacturing. Our findings are
consistent with the role of data requirements, and participants expressed an
expectation related to explainability: that AI should suggest rather than
decide. This preference for human review appeared across functions and
favoured use cases with a clear review point.

The use-case categories group use cases by the kind of support they provide
rather than by the organisational function in which they arose. This allowed
related use cases from different functions to be examined together rather
than as separate ideas. It also provided a common structure for discussing
adoption needs and priorities with company stakeholders.

\subsection{Research and Practical Implications}

For researchers, the proposed approach provides a field study process for
examining the early stages of organisational AI adoption. Researchers can
adapt the stages of preparation, cross-functional data collection, iterative
analysis, stakeholder validation, and reporting to other organisational
settings. Applying the approach in additional organisations could support
comparisons of adoption needs, use case categories, and implementation
conditions across sectors.

For practitioners, the case indicates how such an engagement can be set up. An
organisation can engage an AI researcher for a fixed period and assign a focal
person to support the work. Together, they can examine daily routines,
business processes, data, and system dependencies to identify feasible AI use
cases. The case also suggests that organisations should expect their first
feasible use cases to be those built on information that already exists, and
should treat data integration as a prerequisite for forecasting and predictive
applications. The resulting use cases and priorities can guide
proof-of-concept development, data preparation, system integration, and
further investment.

\textbf{Limitations.} The study examined a single engagement up to the
delivery of the industrial report. It did not follow whether the identified
use cases were implemented afterwards, and it did not compare the approach
with other ways the organisation could have identified AI opportunities, such
as an external consultancy assessment or an internal initiative. The
proof-of-concept demonstrations illustrated selected use cases but were not
evaluated in daily operations, so the study does not show effects on
productivity, cost, work practices, or long-term organisational impact.

\subsection{Threats to Validity}
\label{sec:threats}

\textit{Construct validity.}
Feasibility was assessed through the completion of the six-week engagement and
the company's review of its outputs. Participants were not asked to evaluate
the approach itself, so the claims in this paper are limited to feasibility
and initial utility. Future studies should collect participant feedback on the
approach and follow the use cases during implementation.

\textit{Internal validity.}
The researchers designed the engagement, collected the data, and identified
and prioritised the use cases, which may have influenced the findings. The
first author conducted the detailed coding, and the themes and
interpretations were discussed with the second author and compared with
meeting notes, observations, internal documents, and post-session summaries.
The industrial representative was also part of the research team, which
provided organisational context but means that the account is partly an
insider account.

\textit{External validity.}
The study was conducted in one mid-sized energy company. Differences in
organisational structure, data access, and management support may lead to
different results elsewhere, and further studies are needed in other sectors.

\textit{Reliability.}
Eight interviews were transcribed and eight were documented through
structured notes. The theme counts are therefore based on the transcribed
sessions only, while all 16 sessions informed the use cases. The use cases
were prioritised qualitatively, without a scoring model; the criteria and
reasoning are reported in Section~\ref{sec:results}, and a structured scoring
procedure would strengthen future applications. Because of the NDA, raw
transcripts and internal documents cannot be shared, and participant roles
cannot be reported by organisational function.

\section{Conclusion and Future Work}
\label{sec:conclusion}

In this paper, we proposed the \textit{Adopt-an-AI-Researcher} approach as an entry point for organisations that do not know where or how to start with AI adoption. The approach combines researchers' knowledge of AI capabilities with employees' knowledge of organisational processes, data, constraints, and business needs through a fixed-period engagement.

We applied the approach in an exploratory six-week case study in a energy company. The engagement identified organisational AI adoption needs and produced 41 AI use cases, a use case taxonomy, a priority order of use-case categories, and two proof-of-concept demonstrations. The findings provide evidence of the feasibility and initial utility of the approach, but not yet of its effectiveness or organisational impact. The case also showed that data readiness, rather than how often a need was raised, determined which use cases could be started first.

Future research should expand this approach to other organisations, collect participant feedback on its effort and value, compare it with other ways of identifying AI opportunities, apply a scoring procedure for prioritisation, and examine how prioritised use cases progress to deployment and operational evaluation.

\section*{Declaration on Generative AI}
During the preparation of this work, the authors used AI tools to support
drafting and language refinement, including rephrasing, restructuring, and
improving the clarity of text written by the authors. All text produced
with these tools was critically reviewed and edited by the authors; no AI
output was included without editing. The authors take responsibility for the accuracy, originality,and integrity of the publication.

\bibliographystyle{splncs04}
\bibliography{Reference}








\end{document}